\documentstyle[twoside,fleqn,espcrc2,epsf]{article}

\newcommand{\smallpspicture}[1]{\centerline{\setlength\epsfxsize{7.5cm}\epsfbox{#1}}}
\newcommand{\tinypspicture}[1]{\centerline{\setlength\epsfxsize{6.6cm}\epsfbox{#1}}}

\newcommand{\bi}{\begin{itemize}}
  \newcommand{\ei}{\end{itemize}}

\newcommand{\beq}{\begin{equation}}
  \newcommand{\eeq}{\end{equation}}

\newcommand{\tkl}{{{\small\sf T}{$\chi$}{\small\sf L}} }
\newcommand{\tkll}{{{\sf T}{$\chi$}{\sf L}}}

\newcommand{\sesam}{{\small\sf SESAM }}
\newcommand{\sesams}{{\small\sf SESAM}'s }
\newcommand{\sesamc}{{\small\sf SESAM}-collaboration}

\title{Light Quark Physics with Dynamical Wilson
  Fermions\thanks{Talk presented by H.~Hoeber.}}

\author{{\sf SESAM} and \tkll Collaborations: H.~Hoeber$^{\rm b}$,
                                   N.~Eicker$^{\rm a}$, 
                                   U.~Gl\"assner$^{\rm b}$,
                                   S.~G\"usken$^{\rm b}$,
                                   P.~Lacock$^{\rm a}$, 
                                   Th.~Lippert$^{\rm a}$, 
                                   G.~Ritzenh\"ofer$^{\rm b}$,
                                   K.~Schilling$^{\rm a,b}$,
                                   A.~Spitz$^{\rm a}$,
                                   P.~Ueberholz$^{\rm b}$, 
                                   J.~Viehoff$^{\rm b}$; 
                                   L.~Giusti$^{\rm c}$,
                                   G.~Martinelli$^{\rm d}$,
                                   F.~Rapuano$^{\rm d}$. \\[8pt]
{\small  {\rm $^a$}HLRZ c/o Forschungszentrum J\"ulich, D-52425 J\"ulich,
  and DESY, D-22603 Hamburg, Germany,\\
  {\rm $^b$}Physics Department, University of Wuppertal, D-42097
  Wuppertal, Germany,\\
  {\rm $^c$}Scuola Normale Superiore \& INFN, Pisa and {\rm $^c$}INFN,
                                   University La Sapienza, Roma, Italy.}}
\begin{document}
%%%%%%%%%%%%%%%%%%%%%%%%%%%%%%%%%%%%%%%%%%%%%%%%%%%%%%%%%%%%%%%%%%
\begin{abstract}
We present results for spectroscopy, quark masses and decay constants
obtained from \sesams and \tkll's large statistics simulations of QCD
with two dynamical Wilson fermions.
\end{abstract}
\maketitle
%%%%%%%%%%%%%%%%%%%%%%%%%%%%%%%%%%%%%%%%%%%%%%%%%%%%%%%%%%%%%%%%%%
\section{Introduction}
%%%%%%%%%%%%%%%%%%%%%%%%%%%%%%%%%%%%%%%%%%%%%%%%%%%%%%%%%%%%%%%%%%
The generation of gauge configurations with two dynamical Wilson fermions,
started by the \sesam-collaboration in spring '95 and expanded by the
formation of the \tkl-collaboration in '96, is now well under way. The
primary goal, the generation of a set of configurations 
comprising ${\cal O}(5000)$ trajectories at three sea-quark masses
with $m_\pi/m_\rho = 0.84, 0.76, 0.69$ at $\beta = 5.6$ on
$16^3 \times 32$ lattices was achieved well on time at the end of
'96. Given the rather small box-size of this simulation the aim
of the \tkll-collaboration is to study finite size effects by
simulating at the lightest of \sesams mass-values but on a $24^3
\times 40$ lattice. In addition, we are pushing to even smaller
quark masses, hopefully close to $m_\pi/m_\rho \simeq 0.5$. Last but
not least, in an attempt to clarify issues surrounding the approach to
the chiral limit, we have added a fourth sea-quark value to the $16^3
\times 32$ simulation and the data acquisition for this mass has just
been completed.  
\par
In this short note, I summarize some of the main findings of our
analysis of two-point correlators on these gauge configurations with
the emphasis on the methods we have applied; results are
preliminary and are to be updated soon, when data at the additional
sea-quark mass at $m_\pi/m_\rho = 0.81$ will be included in the
analysis. This note is probably best read in conjunction with
S.~G\"usken's plenary talk writeup \cite{stephan} summarising \sesams
attempt to identify sea-quark effects.
%%%%%%%%%%%%%%%%%%%%%%%%%%%%%%%%%%%%%%%%%%%%%%%%%%%%%%%%%%%%%%%%%%
\section{Results}
\begin{figure}
\vspace{-1cm}
\begin{center}
\centerline{
\smallpspicture{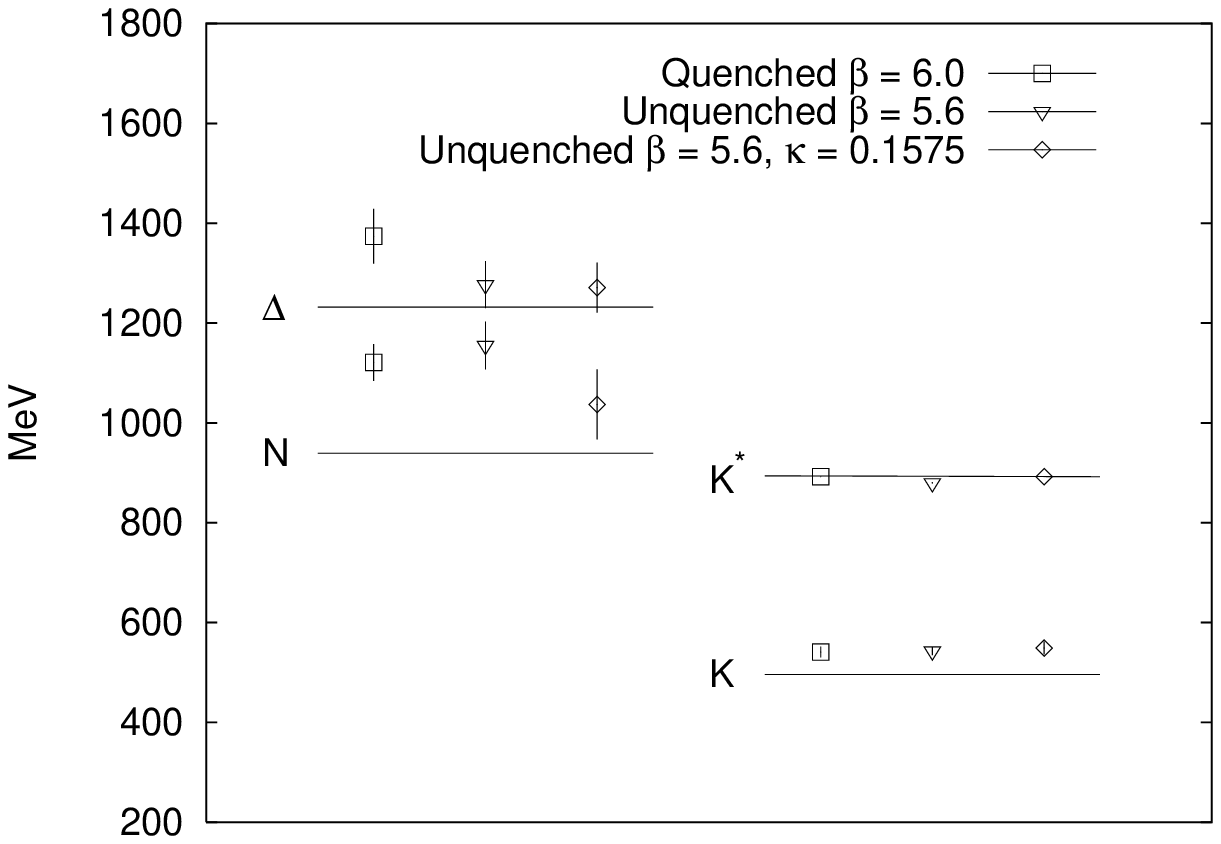}
}
\vspace{-1cm}
\caption{\label{spectrum}Spectrum results.}
\end{center}
\vspace{-1cm}
\begin{center}
\smallpspicture{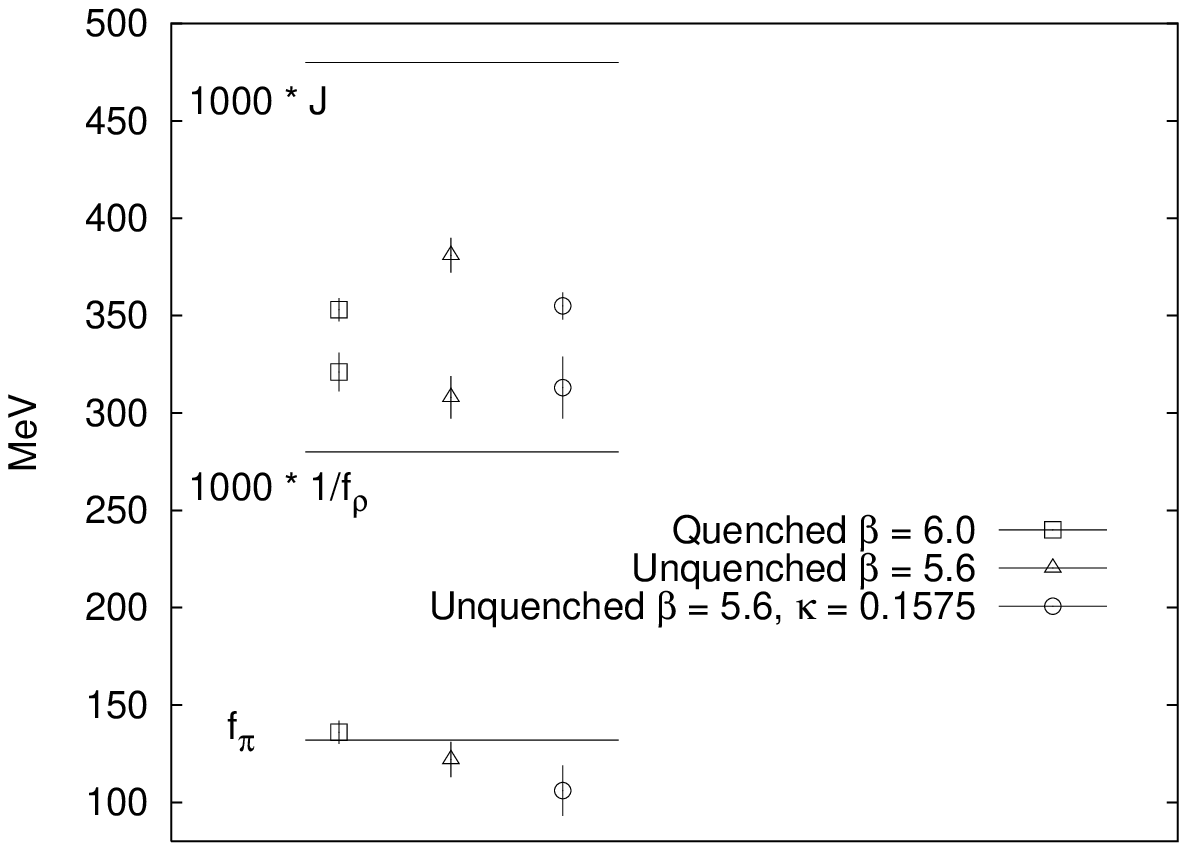}
\end{center}
\vspace{-1.5cm}
\caption{\label{decay}Decay constants and J-parameter.}
\vspace{-0.2cm}
\begin{center}
\smallpspicture{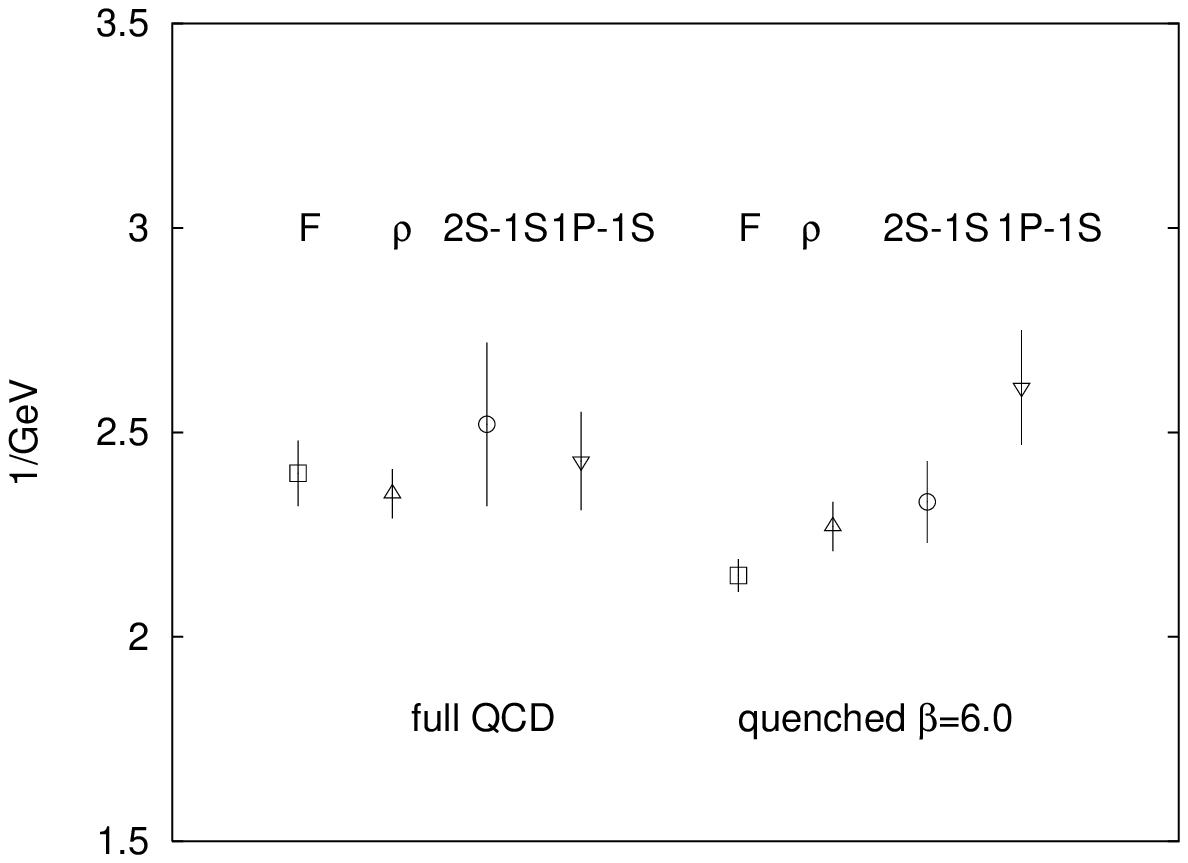}
\end{center}
\vspace{-1.4cm}
\caption{\label{lat}Lattice spacings from different observables ($F$
  denotes the value obtained from the static potential at $r_0$; the
  energy level splittings are obtained from lattice NRQCD \cite{achim}.).}
\end{figure}

%%%%%%%%%%%%%%%%%%%%%%%%%%%%%%%%%%%%%%%%%%%%%%%%%%%%%%%%%%%%%%%%%%
{\bf Error analysis.} One of the first of \sesams results
\cite{sesamauto} was to find that, 
given long enough trajectory lengths in the Hybrid Monte Carlo (HMC),
we can obtain clear signals in the autocorrelation functions for
simple observables (Wilson loops, correlators). We find autocorrelation
times which are much lower,
i.e.${\cal O}(25)$ trajectory lengths, than some of the conjectures
to be found in the early HMC papers. The statistical errors presented in this
paper are obtained from an analysis of 200 configurations (per
sea-quark) picked from a total trajectory length in the HMC of 5000. A
blocking analysis is performed and the quoted 
errors are always from a blocksize of 6, where we find errors to run
into plateaus. I believe this is the first time trajectory lengths in
a HMC are long enough to present a reliable error estimate! Obviously,
this is a vital step in any attempt to find signals of unquenching,
particularly in the area of spectrum calculations since quenched
spectra are (i) in agreement with experiment within ${\cal O}(10 \%)$ 
(see e.~g. \cite{speclat97}) and (ii) can be measured to
ever higher precision.
\\
%%%%%%%%%%%%%%%%%%%%%%%
%%\begin{figure}
%%\begin{center}
%%\centerline{
%%\smallpspicture{geom.eps}
%%}
%%\end{center}
%%\vspace{-7cm}
%%\caption{\label{geom}}
%%\vspace{-1cm}
%%\end{figure}
%%%%%%%%%%%%%%%%%%%%%%%
{\bf Analysis with two degenerate light quarks.} 
Next, I would like to discuss some issues raised recently by \sesam
concerning the analysis of dynamical configurations with 2 degenerate
light Wilson fermions \cite{sesamquarks}. The new feature of our
analysis is the use of spectrum data with unequal valence and sea
quark masses, in particular to determine quantities with strange quark
content. We find our data to be well parameterized by a linear ansatz
both in the sea and valence quark masses (see figure 5 in
\cite{stephan}). We have applied this method to the determination of
the strange quark mass \cite{sesamquarks}, which, so far
\cite{guptaquarks}, has only been obtained in a sea of strange
quarks. The effect is a significant increase in the two-flavour
strange quark mass, whereas our dynamical light quark masses are much
lower than the quenched ones. A surprising feature in the analysis is
obtained when we attempt to study the light quark masses at fixed sea
quark mass (this has been termed ``partial quenching''). We find that
valence quark masses need to be tuned to negative values to make the
pseudoscalar masses vanish. In other words: the critical kappa-values
at fixed sea quark lie below the true critical kappa of pions with
equal strange and valence quark content. We have noted
\cite{sesamquarks} that light quark masses measured with respect to
the partially quenched critical kappa values are much higher,
practically in agreement with the values found in the quenched
theory. 
\\
{\bf Spectrum Results.} The spectrum and decay constant results are
displayed in figures \ref{spectrum} and \ref{decay}. We compare the
dynamical results to (i) a quenched simulation we have performed at
$\beta = 6.0$; the two simulations are at comparable lattice
spacings. To mimic the situation of the dynamical simulation, the
quenched chiral extrapolations were performed linearly; (ii) the result
one obtains (partially quenched) by working at a fixed sea-quark mass
of $\kappa = 0.1575$.  
%%%%%%%%%%%%%%%%%%%%%%%
%%%%%%%%%%%%%%%%%%%%%%%
Studying observables with only light quark content, we see some
improvement for the $\Delta$, the decay constants and the $J$
parameter, albeit within large errors. The nucleon, however, seems to
be unimpressed by the presence of sea quarks. In the strange quark
sector the familiar problem from quenched QCD is still present; the
$K$ and the $K^*$ cannot both be matched simultaneously to
experiment. 
\par
\begin{figure}
\begin{center}
\smallpspicture{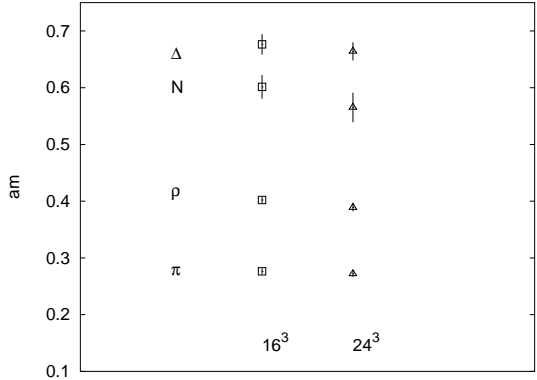}
\end{center}
\vspace{-1.6cm}
\caption{\label{finvol}Finite volume effects in the spectrum data.}
\vspace{-0.7cm}
\end{figure}
\begin{figure}
\begin{center}
\tinypspicture{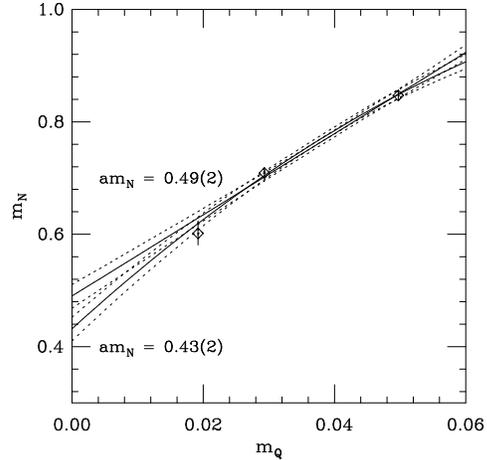}
\end{center}
\vspace{-1.7cm}
\caption{\label{sigmatest}Linear fit of the nucleon data and
  quadratic fit constrained by the $\sigma$ term.}
\vspace{-0.3cm}
\end{figure}

In figure \ref{lat} I show the lattice spacings obtained from four
different observables in light and heavy quark physics. Much better
agreement seems to be obtained in the full theory.
\\
{\bf Finite volume errors.} It was pointed out last year by
S.~Gottlieb \cite{specgottlieb} that little is known so far about the size of finite
volume errors in dynamical Wilson fermion simulations. Figure
\ref{finvol} shows the spectrum data from
the \sesam and \tkl collaborations at $m_\pi/m_\rho = 0.69$ obtained
on lattices of $16^3$ and $24^3$. While the errors are yet too large
for definitive statements there is a general downward trend of the
order of $2, 3$ and $5 \%$ respectively for pseudoscalar, vector and
nucleon masses.  
%%%%%%%%%%%%%%%%%%%%%%%
%%%%%%%%%%%%%%%%%%%%%%%
\\
{\bf Quadratic contributions.} Given that we have been using only
three values of the sea-quark mass in our present analysis we have been restricted to
linear chiral extrapolations. Particularly for the vector but also
for the nucleon, the data seem to exhibit some downward curvature
towards lighter sea-quark masses. Even with only three sea-quark masses
at our disposition there is a nice consistency check for the curvature
of the nucleon due to the Feynman-Hellman theorem which combines the
spectrum analysis with our calculation of the $\pi$-nucleon sigma term
\cite{stephan,jochen}. Given that
\beq
\sigma = m_q \left< P| ( u \bar u + d \bar d)|P \right> = m_q
{\partial m_P \over \partial m_q} \, ,
\label{constrain}
\eeq
we can use our results for $\sigma$ to constrain the linear and
quadratic terms of the proton mass. The result of this test is shown
in figure \ref{sigmatest} where the linear fit to the data is also
plotted. The nucleon data are very well described by
eq.~\ref{constrain}. The effect is to lower the mass of the nucleon in
the chiral limit by ${\cal O}(10 \%)$. It is a very pleasing result to
find the spectrum data match up so well with the flavour-singlet
operator calculation.
%%%%%%%%%%%%%%%%%%%%%%%
%%%%%%%%%%%%%%%%%%%%%%%

%%%%%%%%%%%%%%%%%%%%%%%%%%%%%%%%%%%%%%%%%%%%%%%%%%%%%%%%%%%%%%%%%%
\section{Conclusions}
%%%%%%%%%%%%%%%%%%%%%%%%%%%%%%%%%%%%%%%%%%%%%%%%%%%%%%%%%%%%%%%%%%
We are currently in the middle of the analysis of the spectrum data of
\sesams and \tkl's large statistics simulation of QCD with 2 dynamical
flavours. We now have good control over statistical errors and have
developed a new method to analyse particles with strange quark content
in a sea of light quarks. We are pushing to lighter quark masses,
$m_\pi/m_\rho \approx 0.5$ and to larger lattices, enabling a badly
needed finite volume study. Whereas the quark masses are found to be
rather sensitive to the inclusion of sea-quarks a clear sign of
unquenching is still missing in the spectrum data.
%%%%%%%%%%%%%%%%%%%%%%%%%%%%%%%%%%%%%%%%%%%%%%%%%%%%%%%%%%%%%%%%%%

\end{document}